\documentclass[twocolumn,english,twocolumn,english,aps,prb,floats,showpacs]{revtex4}
\usepackage[T1]{fontenc}
\usepackage[latin9]{inputenc}
\usepackage{bm}
\usepackage{amsmath}
\usepackage{amssymb}
\usepackage{graphicx}
\usepackage{esint}

\makeatletter
\@ifundefined{textcolor}{}
{%
 \definecolor{BLACK}{gray}{0}
 \definecolor{WHITE}{gray}{1}
 \definecolor{RED}{rgb}{1,0,0}
 \definecolor{GREEN}{rgb}{0,1,0}
 \definecolor{BLUE}{rgb}{0,0,1}
 \definecolor{CYAN}{cmyk}{1,0,0,0}
 \definecolor{MAGENTA}{cmyk}{0,1,0,0}
 \definecolor{YELLOW}{cmyk}{0,0,1,0}
}

\usepackage{babel}

\usepackage{babel}

\usepackage{babel}

\makeatother

\usepackage{babel}
\begin{document}

\title{Quantum decay of the persistent current in a Josephson junction ring}

\author{D. A. Garanin and E. M. Chudnovsky}

\affiliation{Physics Department, Lehman College, The City University of New York,
250 Bedford Park Boulevard West, Bronx, NY 10468-1589, U.S.A.}

\date{\today}
\begin{abstract}
We study the persistent current in a ring consisting of $N\gg1$ Josephson
junctions threaded by the magnetic flux. When the dynamics of the
ring is dominated by the capacitances of the superconducting islands
the system is equivalent to the $xy$ spin system in 1+1 dimensions
at the effective temperature $T^{*}=\sqrt{2JU}$, with $J$ being
the Josephson energy of the junction and $U$ being the charging energy
of the superconducting island. The numerical problem is challenging
due to the absence of thermodynamic limit and slow dynamics of the
Kosterlitz-Thouless transition. It is investigated on lattices containing
up to one million sites. At $T^{*}\ll J$ the quantum phase slips
are frozen. The low-$T^{*}$ dependence of the persistent current
computed numerically agrees quantitatively with the analytical formula
provided by the spin-wave approximation. The high-$T^{*}$ behavior
depends strongly on the magnetic flux and on the number of superconducting
islands $N$. Depending on the flux, the persistent current gets destroyed
by the phase slips and/or by the superconductor-insulator transition
on increasing $T^{*}$.
\end{abstract}


\maketitle

\section{Introduction}

Microscopic chains of Josephson junctions have been at the forefront
of research on quantum phase transitions \cite{Girvin,Sachdev} and
quantum circuitry \cite{Ioffe,Gladchenko,Manucharyan}. They provide
a testing field for two fundamental physical effects: Quantum phase
slips \cite{Zaikin,Golubev,Rastelli} and superconductor-insulator
transition \cite{Bradley,Korshunov,Chow,Mooij-2015}. Persistent currents
in small metallic rings have been studied theoretically and experimentally
since 1960s \cite{Oreg} while the studies of microscopic Josephson
junction rings are more contemporary. They are rapidly advancing due
to the progress in manufacturing of the nanostructures \cite{Mooij-2015}.

Analytical research on Josephson junction rings focused on two limits:
When the dynamics of the ring is dominated by the capacitances of
the junctions \cite{Choi-1993} and when the dynamics is dominated
by the capacitances of the superconducting islands \cite{Matveev}.
Also the mixed situation with both capacitances has been studied.
The persistent currents were computed numerically for the rings containing
up to 40 superconducting islands \cite{Lee-2003}, as well as analytically
using the effective low-energy description \cite{Rastelli}. Quantum
phase slips in a Josephson junction chain have been studied experimentally
\cite{Pop} and the good agreement with theoretical concepts \cite{Matveev}
has been demonstrated.

\begin{figure}[htbp!]
\centering \includegraphics[width=8cm]{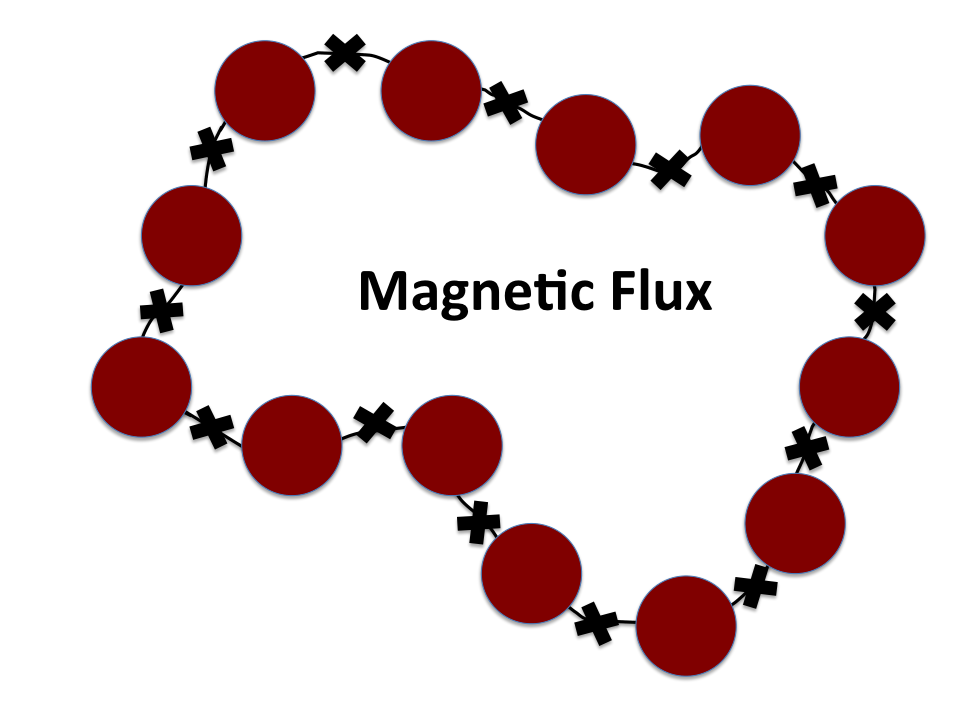} \caption{Color online: A ring of $N$ Josephson junctions (crosses) made of
superconducting grains (circles), threaded by the magnetic flux.}

\label{Fig:JJring}
\end{figure}

Our interest to the problem has been motivated by the strong size
effect observed in the previous numerical studies of the persistent
currents in relatively small Josephson junction rings, Fig. \ref{Fig:JJring}.
It demonstrated the necessity to study longer chains with a large
number of superconducting islands $N$. The case when the dynamics
of the ring is dominated by the capacitances of the islands, that
is considered here, permits such large-$N$ analysis because the quantum
$1d$ problem of the ring maps onto a classical $2d$ $xy$ spin model
at a finite temperature that is well suited for large-scale Monte
Carlo (MC) studies.

The equilibrium persistent current in a Josephson junction ring threaded
by the magnetic flux depends on the value of the flux. It may be destroyed
by temperature or by quantum fluctuations. The measure of the strength
of quantum fluctuations is the ratio of the charging energy of the
superconducting island, $U$, and the Josephson coupling between the
islands, $J$. Exact mapping of the problem onto the $2d$ $xy$ spin
problem allows one to express the effect of quantum fluctuations at
zero temperature in terms of the effective ``quantum'' temperature
$T^{*}=\sqrt{2JU}$. The Kosterlitz-Thouless (KT) temperature, $T_{KT}\sim J$,
at which the quasi-long-range order in the $2d$ model is destroyed
by the unbinding of vortex-antivortex pairs provides the critical
value of $U$ that results in the superconductor-insulator transition.

When the flux is different from $(n+1/2)\Phi_{0}$, where $n$ is
an integer and $\Phi_{0}$ is the flux quantum, the equilibrium persistent
current has a non-zero value as long as quantum fluctuations are weak
$T^{*}<T_{KT}$. With increasing quantum fluctuations one should expect
the persistent current to become zero at $T^{*}=T_{KT}$. The situation
is more complicated in the half-fluxon case, $\Phi=(n+1/2)\Phi_{0}$,
when quantum phase slips provided by vortices of the $2d$ $xy$ classical
model connect, via quantum tunneling, the states with persistent currents
running in opposite directions. In this case the MC routine elucidates
the fact that while the persistent current is theoretically metastable
at any non-zero $U$, the phase slips are frozen at $T^{*}\ll T_{KT}$,
making small Josephson junction rings to resemble superparamagnetic
particles.

The numerical solution of the problem is challenging in comparison
with the typical problems of magnetic phase transitions for two reasons.
First, the persistent current is a mesoscopic quantity that becomes
small for a large system size, $I\propto1/N$. This is why the increase
in the system size does not lead to a significant improvement of the
results via self-averaging. At high quantum temperatures the root-mean-square
fluctuations of the persistent current, $\sqrt{\left\langle I^{2}\right\rangle }\propto1/N$,
decrease with the size but so does the current itself. The second
reason is that the $2d$ $xy$ model does not exhibit a dramatic change
in the spin-spin correlation function due to the unbinding of vortex-antivortex
pairs. The (pseudo)dynamics of this process is extremely slow. It
requires a large number of updates and computer runs.

The paper is structured as follows. The theory is given in Section
\ref{Theory}. The model of a Josephson junction ring whose dynamics
is dominated by capacitances of superconducting islands is formulated
in Section \ref{Sec:Model}. The equivalence of the model to the $xy$
spin model in 1+1 dimensions at finite temperature is reviewed in
Section \ref{Sec:Dynamics}. Persistent current and its $T^{*}$-dependence
expected from the theory are discussed in Section \ref{Sec:Current}.
Numerical results are presented in Section \ref{Sec:Numerics}. The
numerical method is described in Section \ref{Sec:Numerical_Method}.
Formation of vortices responsible for the phase slips and for quantum
KT transition is discussed in Section \ref{Sec:KT}. In Section \ref{Sec:Current decay}
we present numerical results on the destruction of the persistent
current by quantum fluctuations. The energy barrier for the phase
sleep is analyzed in Section \ref{Sec:Analysis}. Section \ref{Sec:Discussion}
contains some final remarks and suggestions for experiment.

\section{Theory}

\label{Theory}

\subsection{Ring made of Josephson junctions}

\label{Sec:Model}

Consider a Josephson junction ring depicted in Fig. (\ref{Fig:JJring}).
Let $\theta_{j}$ be the phase of the superconducting order parameter
$\Psi=|\Psi|\exp(i\theta)$ at the $j$-th superconducting island.
The Josephson energy of the ring is then given by \cite{Tinkham}
\begin{equation}
E_{J}=J\sum_{i}\left[1-\cos\left(\theta_{i+1}-\theta_{i}+\frac{2\pi}{\Phi_{0}}\int_{i}^{i+1}{\bf A}\cdot d{\bf l}\right)\right],\label{H-2dRing}
\end{equation}
where the vector potential ${\bf A}$ is due to the magnetic flux
$\Phi$ piercing the ring.

The summation of phases along the closed loop of the ring gives
\begin{equation}
\sum_{i}\left(\theta_{i+1}-\theta_{i}+\frac{2\pi}{\Phi_{0}}\int_{i}^{i+1}{\bf A}\cdot d{\bf l}\right)=2\pi\left(\phi+m\right),\label{sum}
\end{equation}
where $\phi\equiv\Phi/\Phi_{0}$, $\Phi_{0}=h/(2e)$ is the flux quantum,
and $m$ is an arbitrary integer. This can be derived by computing
the flux through the ring as $\Phi=\oint{\bf A}\cdot d{\bf l}$ and
noticing that the superconducting current ${\bf j}_{s}=\frac{e\hbar}{m}|\Psi|^{2}\left(\bm{\nabla}\theta-\frac{2\pi}{\Phi_{0}}{\bf A}\right)$
is zero inside the islands since we only have Josephson currents in
the system.

\begin{figure}[htbp!]
\centering \includegraphics[width=8cm]{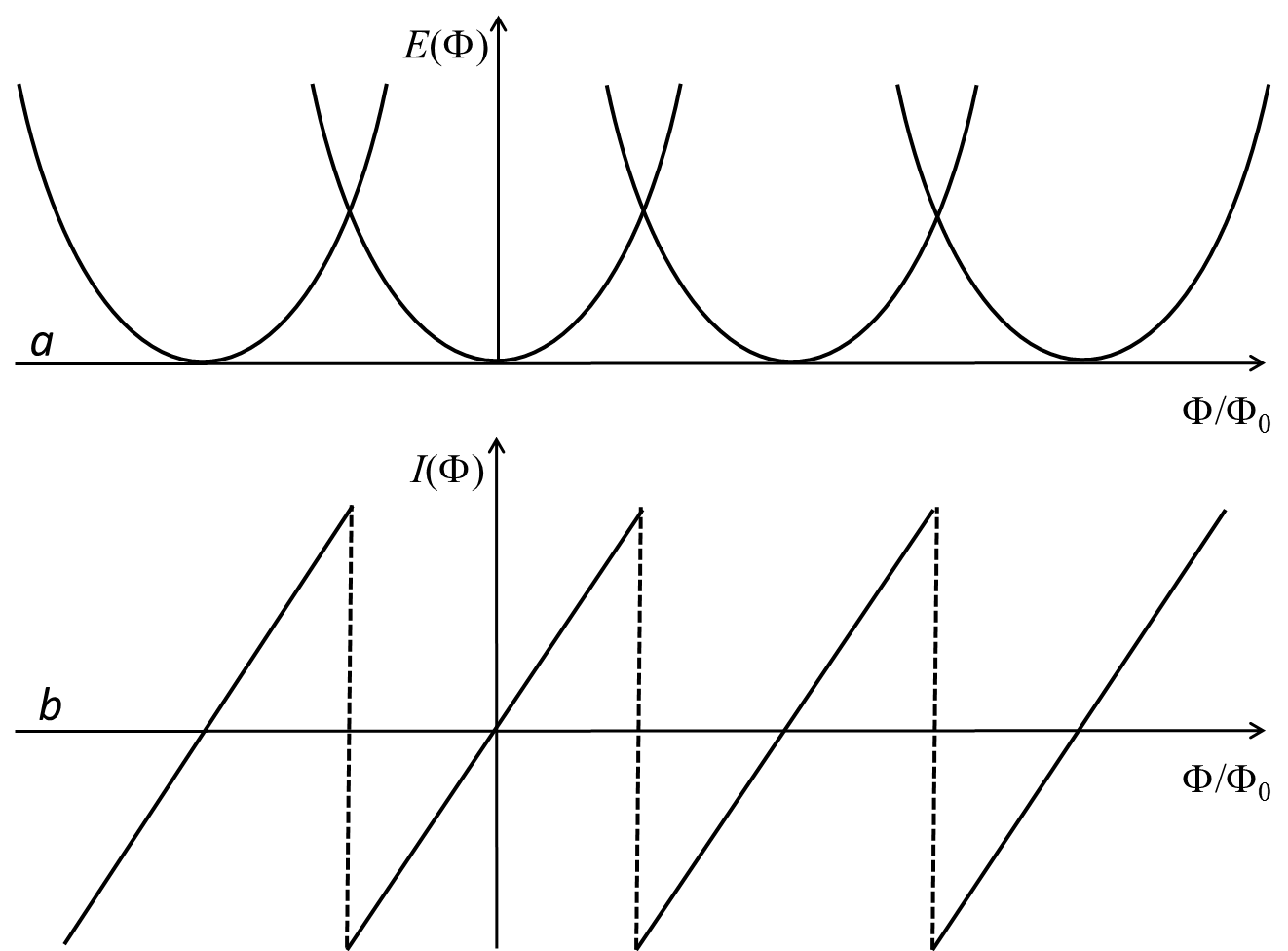} \caption{($a$) $m$-branches of the ground-state energy $E_{J}^{(0)}$. ($b$)
$m$-branches of the persistent current. }

\label{Fig:m-branches}
\end{figure}

Eq. (\ref{sum}) allows one to write
\begin{equation}
E_{J}=J\sum_{i}\left[1-\cos\left(\tilde{\theta}_{i+1}-\tilde{\theta}_{i}+\frac{2\pi\left(\phi+m\right)}{N}\right)\right],\label{eq:EJ_def}
\end{equation}
where the reduced phases $\tilde{\theta}$ are defined in such a way
that the change of $\tilde{\theta}_{i}$ around the ring is zero.
The total change in the original phase $\theta$ accumulated around
the ring is accounted for by the quantum number $m$. These reduced
phases are convenient for the analytical work, whereas the numerical
work uses the original phases. The energy minimum corresponds to all
phases being the same, leading to the ground state
\begin{equation}
E_{J}^{(0)}=NJ\left[1-\cos\left(\frac{2\pi\left(\phi+m\right)}{N}\right)\right]\cong\frac{\left(2\pi\right)^{2}J}{2N}(\phi+m)^{2},\label{GS}
\end{equation}
the last expression being the large-$N$ case. Branches of $E_{J}^{(0)}(\Phi)$
for different values of $m$ are shown in Fig. \ref{Fig:m-branches}$a$.
When $\phi=n+1/2$, with $n$ being an integer, that is for $\Phi=(n+1/2)\Phi_{0}$,
the ground state is degenerate, $E_{J}(n,m)=E_{J}(n,m'=-2n-m-1)$.
This permits quantum tunneling between $E_{J}(n,m)$ and $E_{J}(n,m')$
that removes the degeneracy. For, e.g., $n=0$, that is, when the
flux equals half a fluxon, $\Phi=\Phi_{0}/2$ ($\phi=1/2$), the states
with $m=0$ and $m=-1$, corresponding to different current states
in the ring, have the same energy. Quantum oscillations between such
states have been observed in experiment \cite{Freedman,Mooij}.

\subsection{Dynamics}

\label{Sec:Dynamics}

The dynamics of the model is due the electrical charging of the superconducting
islands by the excess (or lack) of Cooper pairs $n_{i}$ at the $i$-th
site. It is determined by the finite capacitances of the islands and
of the junctions. In this paper we are considering the limit in which
the capacitances of the islands $C$ greatly exceed the capacitances
of the junctions. This situation can be easily achieved in experiment.
For example it will occur when the Josephson junctions are formed
by the weak links between small superconducting grains. In this case
the charging energy is given by
\begin{equation}
E_{C}=\sum_{i}Un_{i}^{2}=\frac{\hbar^{2}}{4U}\sum_{i}\left(\frac{d\theta_{i}}{dt}\right)^{2}\label{EC}
\end{equation}
where $U=(2e)^{2}/(2C)$. The second of Eq.\ (\ref{EC}), which plays
the role of the kinetic energy, is obtained by noticing that $n_{i}$
and $\theta_{i}$ are canonically conjugated variables. If they are
treated quantum-mechanically, one has
\begin{equation}
n_{i}=-i\frac{d}{d\theta_{i}},\qquad i\hbar\frac{d\theta_{i}}{dt}=[\theta_{i},E_{C}]=2iUn_{i}
\end{equation}

Quantum mechanics of the model is formulated in terms of the path
integral
\begin{equation}
I=\prod_{i}\int D\{\theta_{i}(\tau)\}e^{-S_{E}/\hbar}
\end{equation}
where $\tau=it$ and $S_{E}=\int d\tau{\cal L}$ is the Euclidean
action with
\begin{eqnarray}
{\cal L} & = & \frac{\hbar^{2}}{4U}\sum_{i}\left(\frac{d\theta_{i}}{d\tau}\right)^{2}\nonumber \\
 & + & J\sum_{i}\left[1-\cos\left(\theta_{i+1}-\theta_{i}+\frac{2\pi\phi}{N}\right)\right].
\end{eqnarray}
Here the phases $\theta_{i}$ are the original phases, as in Eq. (\ref{H-2dRing}),
not the reduces phases of Eq. (\ref{eq:EJ_def}).

\begin{figure}[htbp!]
\centering \includegraphics[width=8cm]{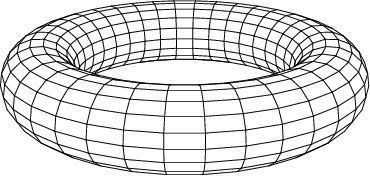} \caption{$(1+1)d$ space-time lattice with periodic boundary conditions used
in numerical work.}

\label{Fig:toroid}
\end{figure}

This quantum model at $T=0$ is equivalent \cite{Girvin} to the statistical
mechanics of the classical model in 1+1 dimensions at a non-zero temperature
$T^{*}=\sqrt{2JU}$, described by the partition function
\begin{equation}
Z=\prod_{i}\int D\{\theta_{i}(\tau)\}e^{-\mathcal{H}_{1+1}/T^{*}}\label{eq:Z_def}
\end{equation}
with
\begin{equation}
\mathcal{H}_{1+1}=-\frac{1}{2}\sum_{\mathbf{rr}'}J_{\mathbf{rr}'}\cos\left(\theta_{\mathbf{r}'}-\theta_{\mathbf{r}}+\phi_{\mathbf{rr}'}\right),\label{eq:H_1+1_theta}
\end{equation}
where $\mathbf{r}$ is a discrete two-dimensional vector $\mathbf{r}=(i,l)$
representing the space-time lattice, while $J_{\mathbf{rr}'}=J$ for
the nearest neighbors and zero otherwise. In the numerical work we
use the $N\times N$ lattice with periodic boundary conditions that
correspond to the surface of a toroid, Fig. \ref{Fig:toroid}. The
circumference of the cross-section of the toroid along the closed
$i$ direction contains $N$ sites corresponding to $N$ superconducting
islands in the Josephson junction ring. The $l$ direction along the
length of the toroid corresponds to the imaginary time. Phase shifts
are given by $\phi_{i,l;i\pm1,l}=\pm2\pi\phi/N$ and $\phi_{i,l;i,l\pm1}=0$.
Notice that, in principle, the periodic boundary condition imposed
on the imaginary time (our closed $x$ direction) introduces a finite
physical temperature into the original quantum problem, $T\sim T^{*}/N$.
At large $N$ the effect of that temperature on the persistent current
can be ignored.

The statistical model presented above can be reformulated in terms
of the two-component classical spin vectors of the $2d$ $xy$ model
at temperature $T^{*}=\sqrt{2JU}$ that describes the strength of
quantum fluctuations. As is known, the $2d$ $xy$ model exhibits
the Kosterlitz-Thouless phase transition at \cite{Tobochnik,Gupta}
$T_{c}\approx0.89J$. For the quantum model this means that on increasing
the charging energy $U$ quantum fluctuations become sufficiently
strong to destroy Josephson currents. It is believed that the corresponding
quantum phase transition results in the state of the ring (sometimes
called the Cooper-pair insulator) in which the islands connected by
Josephson junctions maintain their superconductivity but no Josephson
current can circulate in the ring. The natural way to test this interpretation
of the quantum KT phase transition is to study the $U$-dependence
of the persistent Josephson current in the ring.

\subsection{Persistent current}

\label{Sec:Current}

In accordance with electrodynamics, the persistent current in the
Josepson junction ring is given by
\begin{equation}
I=\frac{d\left\langle E_{J}\right\rangle }{d\Phi}=\frac{1}{\Phi_{0}}\frac{d\left\langle E_{J}\right\rangle }{d\phi},\label{I-GS}
\end{equation}
where averaging is performed over quantum fluctuations with the help
of the statistical model of Eq. (\ref{eq:Z_def}). At $U=0$, when
quantum fluctuations are absent, substitution into this formula of
the ground-state energy $E_{J}^{(0)}(\Phi)$ given by Eq.\ (\ref{GS})
results at large $N$ in $I(\Phi)$ shown in Fig. \ref{Fig:m-branches}$b$.
For, e.g., half a fluxon, $\Phi=\Phi_{0}/2$, the current has the
same absolute value but flows in opposite directions for $m=0$ and
$m=-1$. Any non-zero $U$ permits quantum tunneling between such
current states that has been observed in experiment \cite{Freedman,Mooij}.
Eq. (\ref{eq:EJ_def}) yields
\begin{eqnarray}
\langle E_{J}\rangle & = & NJ\left[1-\left\langle \cos\left(\tilde{\theta}_{i+1;l}-\tilde{\theta}_{i;l}+\frac{2\pi\left(\phi+m_{l}\right)}{N}\right)\right\rangle \right]\nonumber \\
 & = & NJ\left[1-\left\langle \cos\frac{2\pi\left(\phi+m_{l}\right)}{N}\cos\left(\tilde{\theta}_{i+1;l}-\tilde{\theta}_{i;l}\right)\right\rangle \right.\nonumber \\
 & + & \left.\left\langle \sin\frac{2\pi\left(\phi+m_{l}\right)}{N}\sin\left(\tilde{\theta}_{i+1;l}-\tilde{\theta}_{i,l}\right)\right\rangle \right].
\end{eqnarray}
Note that quantum fluctuations involve both, fluctuations of the reduced
phases $\theta_{i,l}$ and the phase slips corresponding to the transitions
between different $m$-numbers. The latter leads to different values
of $m=m_{l}$ at different moments of the discrete imaginary time
$l$.

At small $U$ satisfying $T^{*}=\sqrt{2JU}\ll T_{c}\sim J$ the phase
slips have exponentially small probability. Consequently, at a small
$T^{*}$, if one induces a persistent current by placing the Josephson
junction ring in the magnetic field, the phase slips may not occur
on the time scale of the experiment. In this case the energy simplifies
to
\begin{equation}
\langle E_{J}\rangle=NJ\left[1-\cos\frac{2\pi\left(\phi+m\right)}{N}\left\langle \cos\left(\tilde{\theta}_{i+1,l}-\tilde{\theta}_{i,l}\right)\right\rangle \right],
\end{equation}
since $\left\langle \sin\left(\tilde{\theta}_{i+1,l}-\tilde{\theta}_{i,l}\right)\right\rangle =0$.
The persistent current computed with the help of Eqs. (\ref{I-GS})
becomes
\begin{equation}
I=\frac{2\pi J}{\Phi_{0}}\sin\left(\frac{2\pi\left(\phi+m\right)}{N}\right)\langle\cos(\tilde{\theta}_{i,l}-\tilde{\theta}_{i+1,l})\rangle.
\end{equation}
This expression corresponds to the spin-wave approximation in which
the effect of the magnetic flux and the global phase change $m$ have
been factored out.

We now recall that the statistical mechanics of our model is that
of the $2d$ $xy$ model at $T=T^{*}$, for which the low-temperature
(spin-wave) result is \cite{Tobochnik}:
\begin{equation}
\left\langle \cos\left(\theta_{i}-\theta_{i+\delta}\right)\right\rangle =1-\frac{T^{*}}{4J},
\end{equation}
$\delta$ being the nearest neighbor in any direction. This gives
\begin{eqnarray}
I & = & \frac{2\pi J}{\Phi_{0}}\sin\left(\frac{2\pi\left(\phi+m\right)}{N}\right)\left(1-\frac{T^{*}}{4J}\right)\nonumber \\
 & \cong & \frac{\left(2\pi\right)^{2}J}{N\Phi_{0}}\left(\phi+m\right)\left[1-\left(\frac{U}{8J}\right)^{1/2}\right]\label{eq:I_SWT}
\end{eqnarray}
for the persistent current in the \textquotedbl{}spin-wave\textquotedbl{}
limit. In the last equation we have used $\sin\left[2\pi(\phi+m)/N\right]\cong2\pi(\phi+m)/N$
at large $N$. As we shall see, this formula agrees well with numerical
results.

\begin{figure}
\centering\includegraphics[width=8cm]{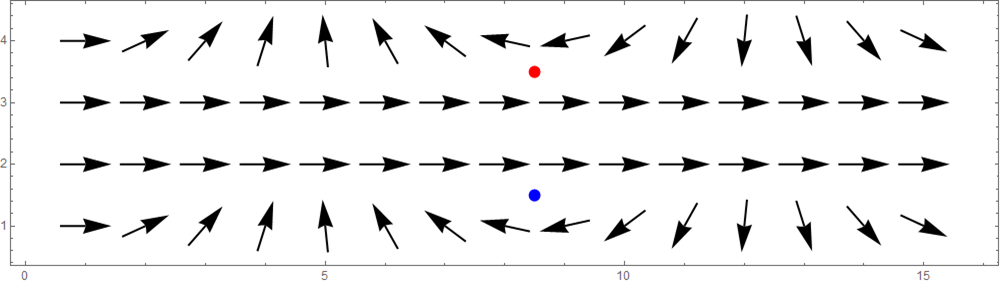}

\caption{Phase slip in $2d$ $xy$ model via creation of the vortex-antivortex
pair.}

\label{Fig:Phase_slip_and_vortices}
\end{figure}

Phase slips require creation of vortex-antivortex pairs, as is illustrated
schematically in Fig. \ref{Fig:Phase_slip_and_vortices}. The phase
of each superconducting island is represented by the spin vector $\mathbf{s}_{\mathbf{r}}=\left(\cos\theta_{\mathbf{r}},\sin\theta_{\mathbf{r}}\right)$.
The vortex (blue dot) is shown at the center of the plaquette for
which, on the way counterclockwise, the spin is rotating by $2\pi$.
For the antivortex (red dot) the spin is rotating by $-2\pi$. Horizontal
direction represents the $1d$ ring, while the vertical direction
represents the imaginary time. In the two central rows $m=0$ while
in the top and bottom rows $m=1$. In the absence of the magnetic
flux, the current in the two central rows is zero. To the contrast,
in the top and the bottom rows there is a current due to the phase
gradient. Slow spatial rotation of the spins in these rows eventually
makes them strongly non-collinear with the spins in the central rows,
which, inevitably, leads to the creation of singularities -- vortices
and antivortices. A closely-bound vortex pair has the energy of order
$J$, thus at $T^{*}\ll J$ the concentration of vortex pairs is exponentially
small. This is why the phase slips can be ignored in the spin-wave
approximation. If the distance between the vortex and the antivortex
in the pair increases, the area where the current is disturbed also
increases. However, an additional energy, that is logarithmic on the
separation, is required to break the pair. According to the established
scenario, the vortex pairs unbind at the temperature $T^{*}=T_{KT}$
of the Kosterlitz-Thouless transition. This would mean unlimited proliferation
of the phase slips and the complete destruction of the persistent
current.

In the half-fluxon case, even below the Kosterlitz-Thouless transition,
vortices provide phase slips that make the persistent current to tunnel
between opposite directions. If one allows sufficient real or computer
time, it makes the initially created persistent current to evolve
to the zero average corresponding to a superposition of clockwise
and counterclockwise currents. It is instructive to compare the Josephson
ring with the Ising system of a finite size. In the thermodynamic
limit, $N\rightarrow\infty$, the $2d$ and $3d$ Ising models possess
the order parameter -- magnetization. This happens, because the transition
from the state with the magnetization looking up to the state with
the magnetization looking down requires the formation of the domain
wall that traverses the system. Its energy scales as the size of the
system $N$. To the contrast, the persistent current is a mesoscopic
quantity, $I\propto1/N$, that disappears in the thermodynamic limit.
If, instead, one deals with the quantity $NI$, the dependence of
the phase-slip barrier on $N$ becomes important. Since only one vortex-antivortex
pair is needed for the phase slip, the corresponding barrier can only
have logarithmic dependence on $N$. Consequently, one should expect
that increasing the system size will not stabilize the persistent
current. Thus, the latter cannot play the role of the order parameter
similar to the magnetization in the Ising model. A more detailed analysis
of the barrier associated with the phase slip will be presented in
Section \ref{Sec:Analysis}.

\section{Numerical results}

\label{Sec:Numerics}

\begin{figure}[htbp!]
\centering \includegraphics[width=8cm]{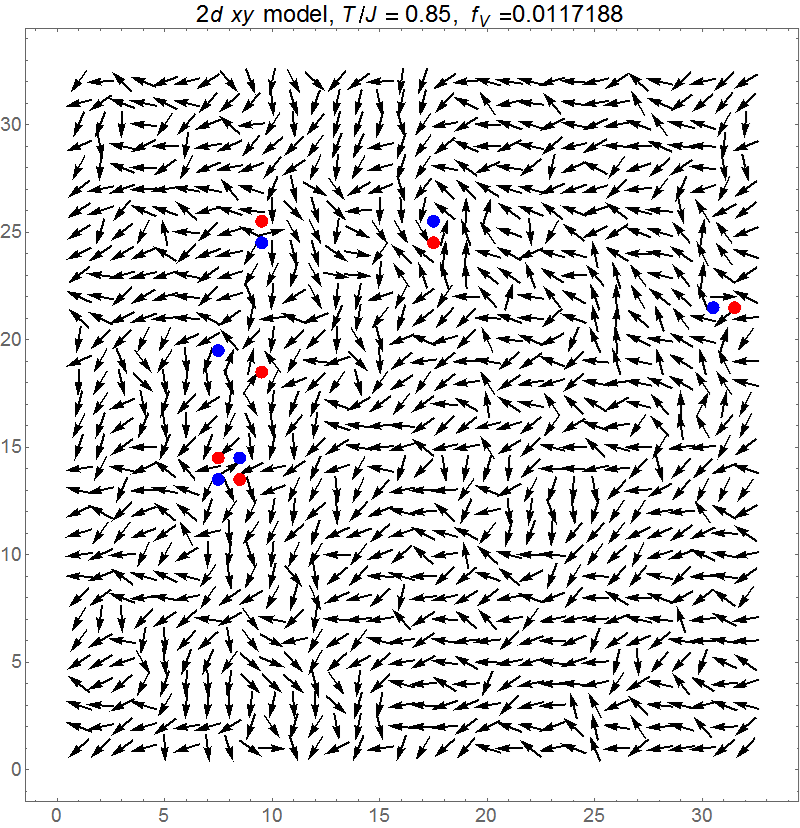}
\caption{Color online: Vortex pairs in the $2d$ $xy$ model providing phase
slips in the quantum Josephson-junction ring.}

\label{Fig:Vortices_T=00003D00003D1}
\end{figure}

\begin{figure}
\centering\includegraphics[width=8cm]{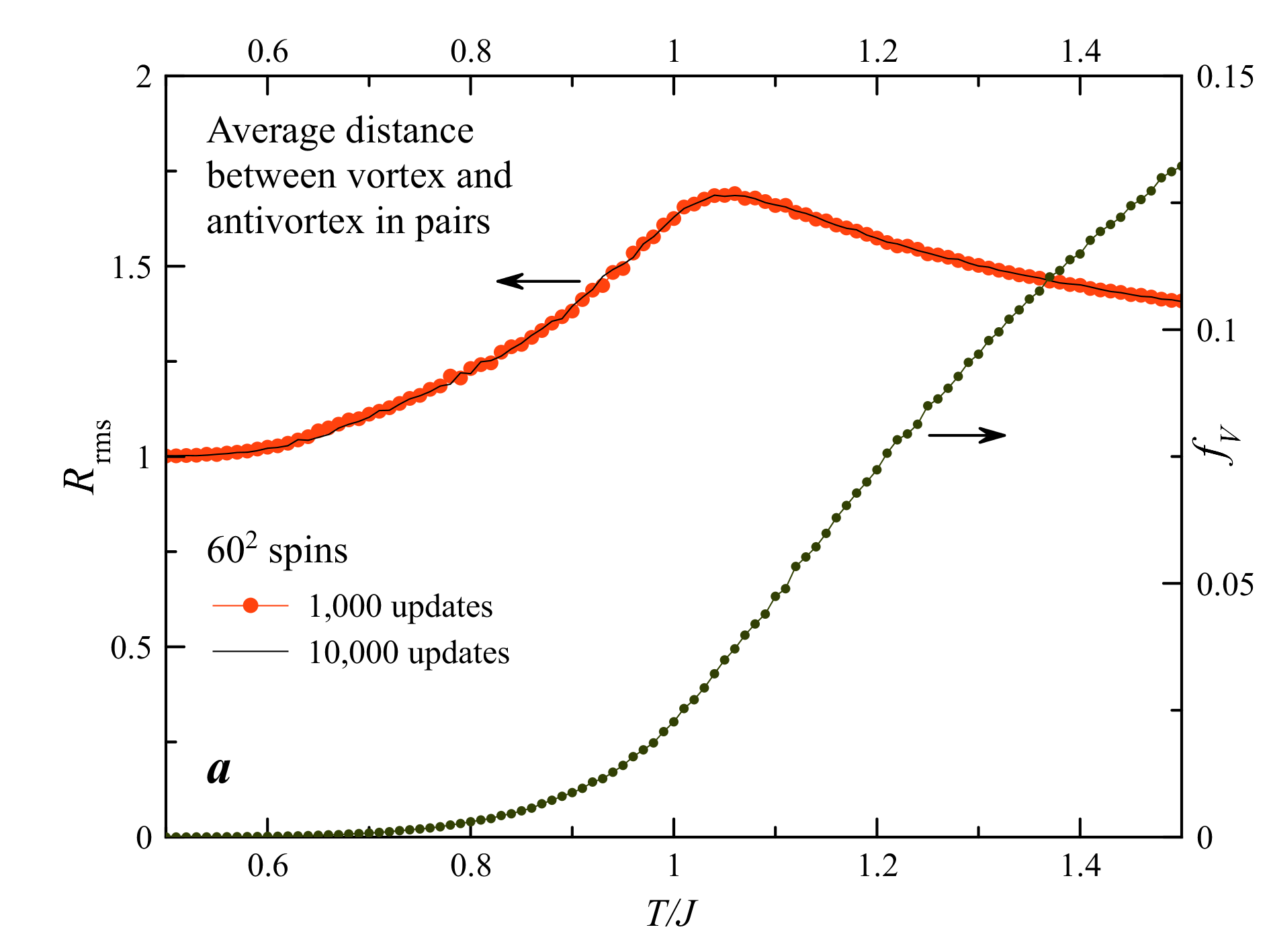}

\includegraphics[width=8cm]{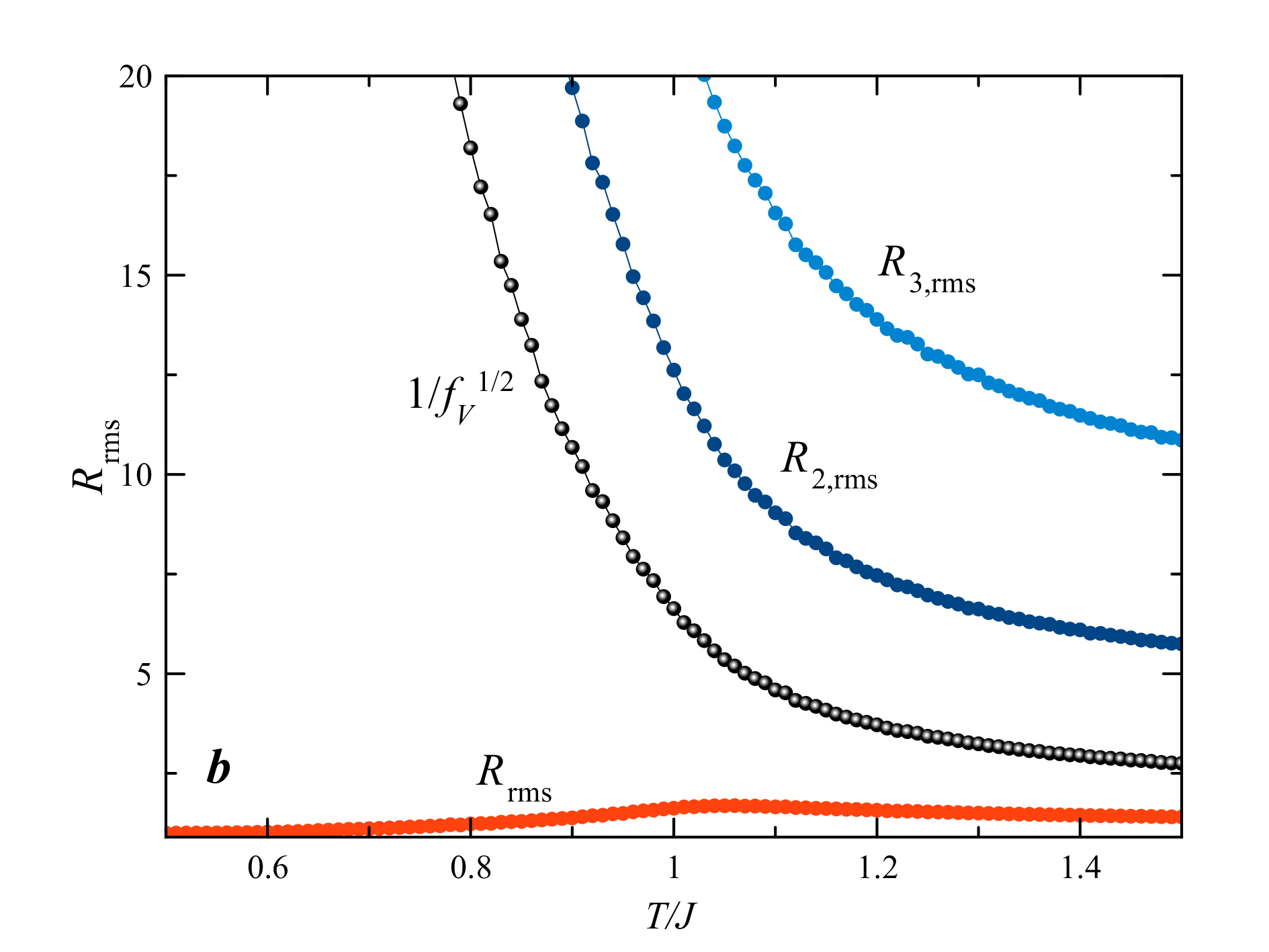}

\caption{Color online: Vorticity $f_{V}$ and the average distance between
vortices and antivortices in pairs vs $T$. ($a$) Vorticity and distance
in the pairs; ($b$) Distances between the nearest ($R_{\mathrm{rms}}$),
next-nearest ($R_{2,\mathrm{rms}}$), and next-next-nearest ($R_{3,\mathrm{rms}}$)
neighbors of a vortex, as well as the average distance between singularities
$1/\sqrt{f_{V}}$. }

\label{Fig:Rrms_vs_T}
\end{figure}

\subsection{Numerical method}

\label{Sec:Numerical_Method}

To solve the problem numerically, it is convenient to rewrite the
effective classical Hamiltonian $\mathcal{H}_{1+1}$ of Eq. (\ref{eq:H_1+1_theta})
in terms of the classical spin vectors $\mathbf{s}_{\mathbf{r}}=\left(\cos\theta_{\mathbf{r}},\sin\theta_{\mathbf{r}}\right)$,
\begin{equation}
\mathcal{H}_{1+1}=-\frac{1}{2}\sum_{\mathbf{rr}'}J_{\mathbf{rr}'}\mathbf{s}_{\mathbf{r}}\cdot\mathbb{R}_{\mathbf{rr}'}\cdot\mathbf{s}_{\mathbf{r}'},\label{eq:H_1+1_spin}
\end{equation}
where
\begin{equation}
\mathbb{R}_{\mathbf{rr}'}=\left(\begin{array}{cc}
\cos\phi_{\mathbf{rr}'} & -\sin\phi_{\mathbf{rr}'}\\
\sin\phi_{\mathbf{rr}'} & \cos\phi_{\mathbf{rr}'}
\end{array}\right)
\end{equation}
and $\phi_{\mathbf{rr}'}$ are defined below Eq. (\ref{eq:H_1+1_theta}).

Averages of physical quantities at temperature $T^{*}$ are computed
by a combination of standard Metropolis updates and over-relaxation.
In a Metropolis update, a spin is rotated by a random trial angle
and the corresponding energy change $\Delta E$ is computed. If $\Delta E<0$,
the rotation is accepted. If $\Delta E>0$, the rotation is accepted
with probability $\exp\left(-\Delta E/T^{*}\right)$. To keep the
acceptance rate not too small and not too large, the trial angles
are kept within the interval that increases with $T^{*}$. The so-called
over-relaxation flips the spin over the effective field $\mathbf{H}_{\mathrm{eff},\mathbf{r}}=-\partial\mathcal{H}_{1+1}/\partial\mathbf{s_{r}}$
according to $\mathbf{s}_{\mathbf{r}}\Rightarrow2\left(\mathbf{s}_{\mathbf{r}}\cdot\mathbf{h}_{\mathrm{eff},\mathbf{r}}\right)\mathbf{s}_{\mathbf{r}}-\mathbf{s}_{\mathbf{r}}$,
where $\mathbf{h}_{\mathrm{eff},\mathbf{r}}=\mathbf{H}_{\mathrm{eff},\mathbf{r}}/H_{\mathrm{eff},\mathbf{r}}$.
Over-relaxation is, in fact, a kind of a conservative pseudo-dynamics.
For each spin, Metropolis update was done with the probability $\alpha$
and the over-relaxation was done with the probability $1-\alpha$.
The constant $\alpha$ plays the role of the damping in our computations.
This routine is performed on all spins sequentially. Updating all
spins means one complete update. It is well-known that for classical
spin systems mixing over-relaxation with the Monte Carlo routine increases
the performance of the numerical method. Metropolis updates cause
dynamics of a diffusive type, making the system to explore its phase
space slowly. Over-relaxation is a fast ballistic process covering
the phase space fast. In our problem, pure Monte Carlo $\left(\alpha=1\right)$
does not result in switching of the current in the half-fluxon case
via transitions that require creation and unbinding of the vortex-antivortex
pairs. Thus, using pure Monte Carlo one can erroneously conclude that
the current is stable. However, when the over-relaxation is dominant
$\left(\alpha\ll1\right)$ the current is switching directions at
the elevated temperatures.

Most of the numerical results were obtained with a two-stage process.
First, 300 updates with $\alpha=1$ were done to equilibrate the system
at a certain temperature. Then, a large number (up to $10^{5}$) of
updates with $\alpha=0.01$ were done for each temperature to explore
the phase space and to allow the switching of the current. The temperature
was typically increased or decreased in small steps ($\Delta T^{*}/J=0.01$)
that provided an additional possibility for equilibration. Finally,
similar runs of the above routine were conducted in parallel on a
multi-core computer and averaging over the runs was performed.

In numerical work we use the original phases of Eq. (\ref{H-2dRing}),
represented by spin vectors. They are more convenient than the reduced
phases $\tilde{\theta}_{i}$ since both phases and quantum numbers
$m$ are fluctuating at elevated temperatures. The average $m$ was
computed as
\begin{equation}
\left\langle m\right\rangle =\frac{1}{2\pi}\left\langle \sum_{i}\left(\theta_{i+1}-\theta_{i}\right)\right\rangle ,\label{sum-1}
\end{equation}
c.f. Eq. (\ref{sum}) that is exponentially close to an integer at
low temperatures.

The software used was Wolfram Mathematica that allows compilation
(including usage of an external C compiler that doubles the speed)
and parallelization. The main operating computer was Dell Precision
T7610 Workstation with two Intel Xeon Processors E5-2680 v2 (10 Core,
2.8GHz each). Mathematica could use 16 cores out of 20. The largest-scale
computations were done for size $256^{2}$ ($N=256$) with 100,000
updates, size $512^{2}$ with 10,000 updates, and size $1000^{2}$
with 3,000 updates for each temperature. The number of runs was about
hundred (see indicated in figures).

\subsection{Vortex pairs in the effective classical $2d$ $xy$ model}

\label{Sec:KT}

The underlying classical $2d$ $xy$ model has been intensively studied
in the past, making the Kosterlitz-Thouless transition the likely
mechanism of the destruction of the persistent current by quantum
fluctuations at the effective quantum temperature $T^{*}=T_{KT}$.
However, in the context of the phase slips that occur in a quantum
$1d$ model, the numerical study of the unbinding of vortices is missing
and will be presented here.

Fig. \ref{Fig:Vortices_T=00003D00003D1} shows thermally excited vortex-antivortex
pairs in a $2d$ $xy$ system of size $32^{2}$ at $T^{*}/J=1$ that
is slightly above the transition temperature at $T_{KT}/J=0.89$.
One can see that vortices and antivortices are still close to each
other, although there is at least one unpaired vortex and one unpaired
antivortex. Fig. \ref{Fig:Rrms_vs_T}$a$ shows the vorticity $f_{V}$,
defined as the number of vortices and antivortices together per plaquette,
for the system of size $60^{2}$. Temperature dependence of the average
root-mean-square distance $R_{\mathrm{rms}}$ between vortex and antivortex
in the pair is shown as well. The vorticity is exponentially small
at low temperatures and it reaches the limiting value 1/3 at $T\rightarrow\infty$.
Near the KT transition the vorticity is still small, $f_{V}\ll1$.
The distance $R_{\mathrm{rms}}=1$ on the left side of the plot corresponds
to the vortex and antivortex in the pair occupying neighboring plaquettes.
With temperature increasing, $R_{\mathrm{rms}}$ increases too, as
expected. Surprisingly, however, it does not become large near the
KT transition, in contrast with the popular narrative of the massive
production of free vortices due to the unbinding of pairs at $T_{KT}$.
The reason for this must be that $f_{V}$ increases with temperature,
so that the average distance between singularities, $r^{*}=a/\sqrt{f_{V}}$,
is decreasing, pressing $R_{\mathrm{rms}}$ down. This is illustrated
in Fig. \ref{Fig:Rrms_vs_T}$b$ showing $R_{\mathrm{rms}}$ together
with the average distances to the first, second, and third nearest
antivortex neighbors for a vortex in a system of size $60^{2}$. Increasing
the system size does not change these results.

\subsection{Destruction of the persistent current by quantum fluctuations}

\label{Sec:Current decay}

\begin{figure}[htbp!]
\centering \includegraphics[width=8cm]{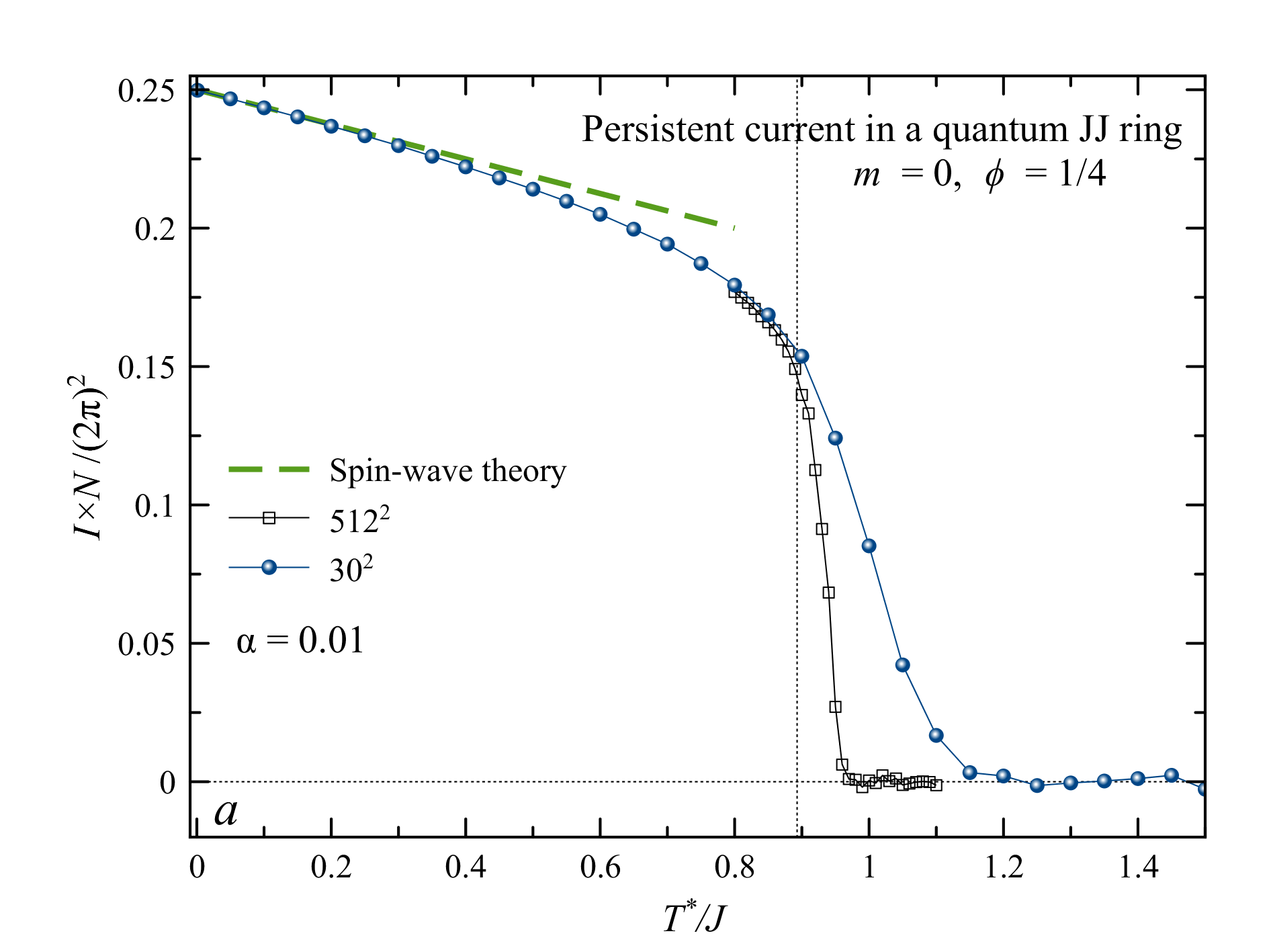} \includegraphics[width=8cm]{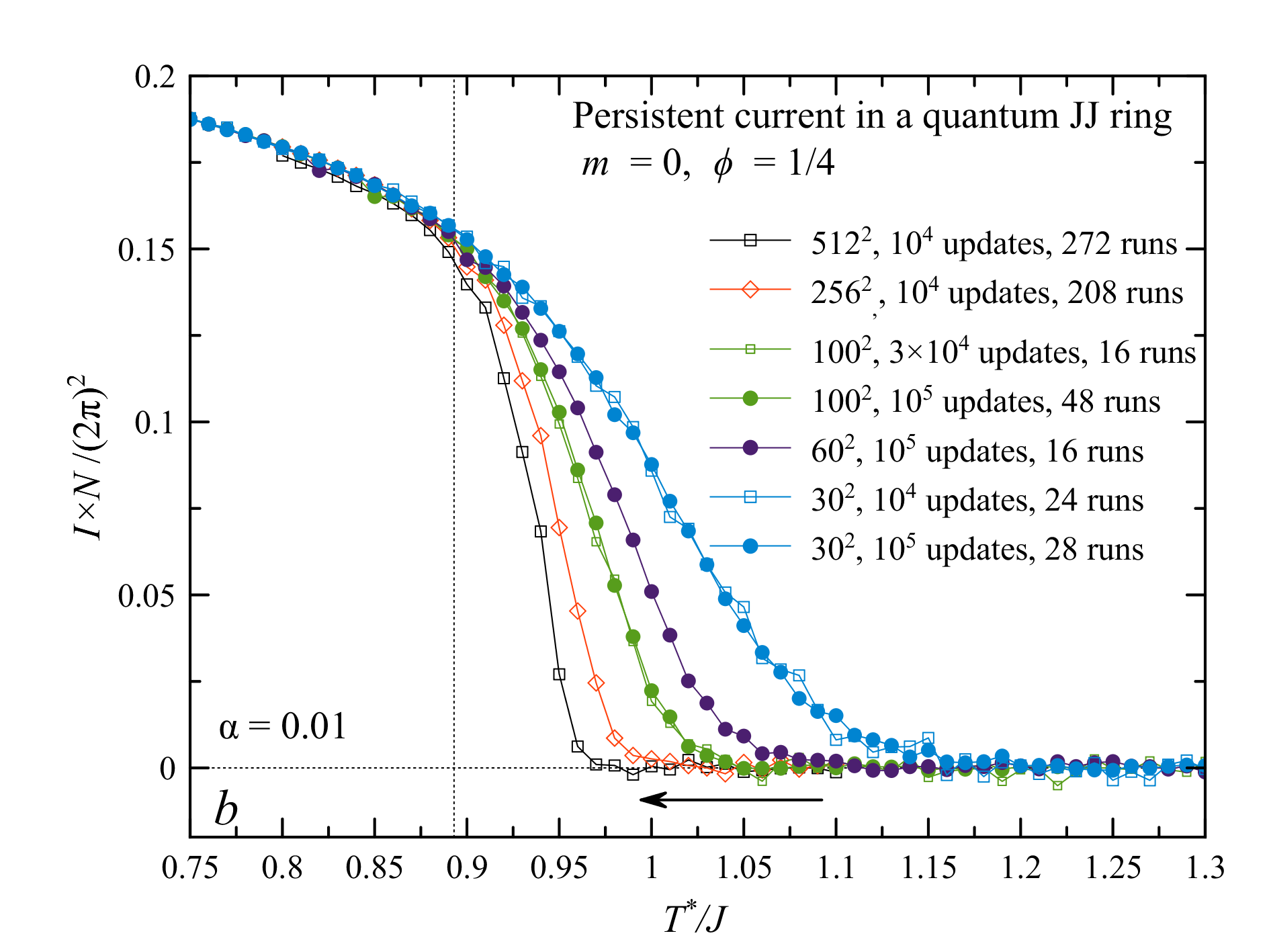}
\caption{Color online: Dependence of the persistent current on the quantum
temperature $T^{*}=\sqrt{JU}$ for different lattice sizes in the
quarter-fluxon case. ($a$) The entire temperature range. The result
of the spin-wave theory is given by Eq. (\ref{eq:I_SWT}). ($b$)
Vicinity of the Kosterlitz-Thouless transition.}

\label{Fig: I_quarter_fluxon}
\end{figure}

Temperature dependence of the persistent current $I$ in the quarter-fluxon
case, $\phi=1/4$, obtained by decreasing $T^{*}$, is shown in Fig.
\ref{Fig: I_quarter_fluxon} for different system sizes $N$. In fact,
$IN$ is plotted. In such representation the curves for different
sizes coincide everywhere except in the region of the KT transition.
The spin-wave theory, Eq. (\ref{eq:I_SWT}), works well in the low-temperature
region. The curves for different sizes diverge at the temperature
very close to $T_{KT}=0.89J$, becoming steeper with increasing the
size. One can project that in the limit $N\rightarrow\infty$ there
will be a jump at $T^{*}=T_{KT}$. However, in this limit the persistent
current itself disappears. Since $I$ becomes small for large sizes,
its fluctuations grow and one has to perform many runs of computations
to average them out. Most of the data have been obtained with $10^{4}$
and $10^{5}$ updates, and the results for different numbers of updates
coincide. This means that in our computations the system is at equilibrium
at any temperature.

\begin{figure}
\centering\includegraphics[width=8cm]{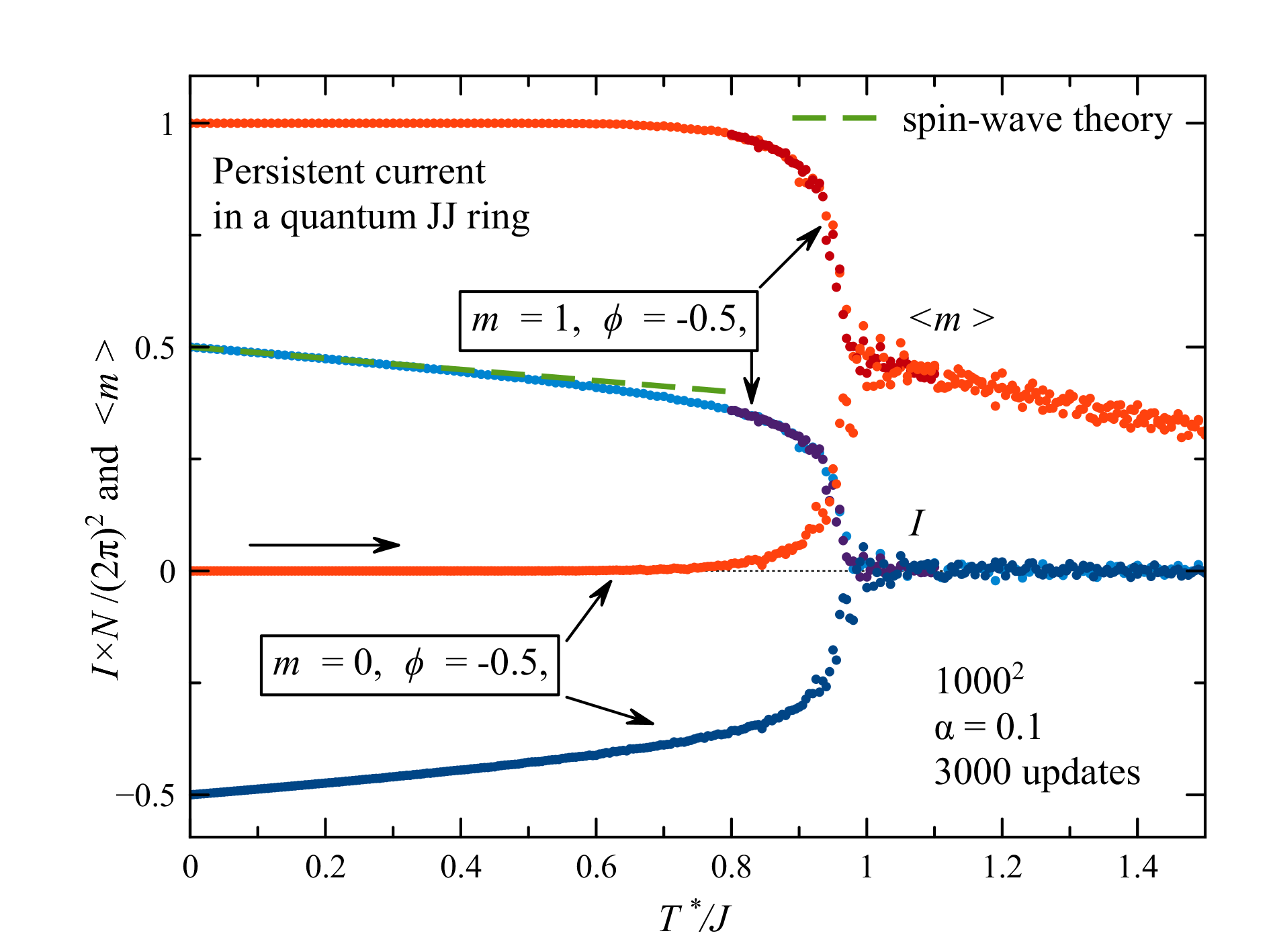}

\caption{Color online: Dependence of the persistent current and the average
phase change $\left\langle m\right\rangle $ on increasing $T^{*}$
in the half-fluxon case for the intermediate damping $(\alpha=0.1)$
and a moderate number of updates. Green dashed line is the spin-wave
result of Eq. (\ref{eq:I_SWT}).}

\label{Fig:I_m_vs_T_Nx=00003D00003DNy=00003D00003D100}
\end{figure}

\begin{figure}
\centering\includegraphics[width=8cm]{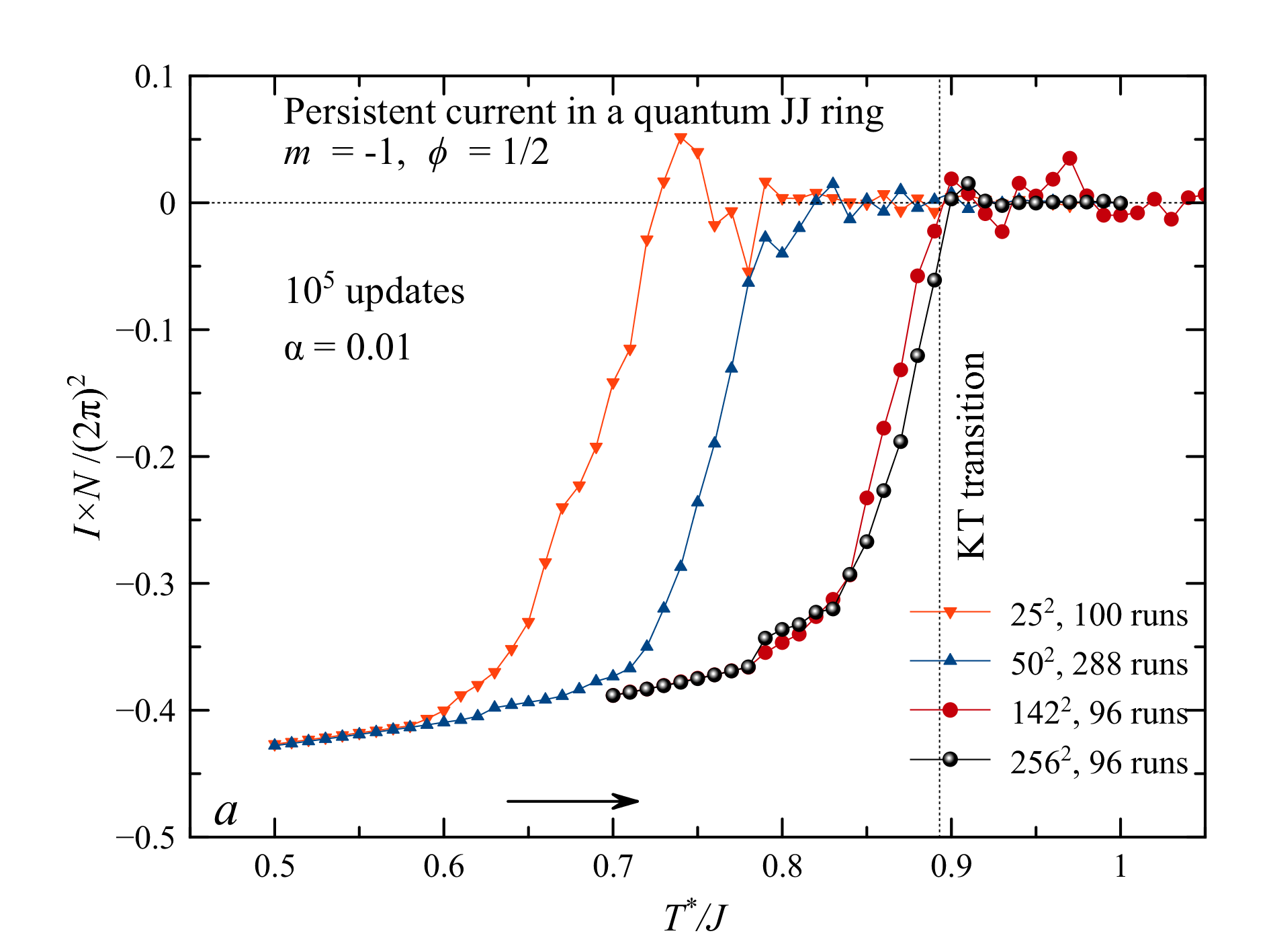}

\includegraphics[width=8cm]{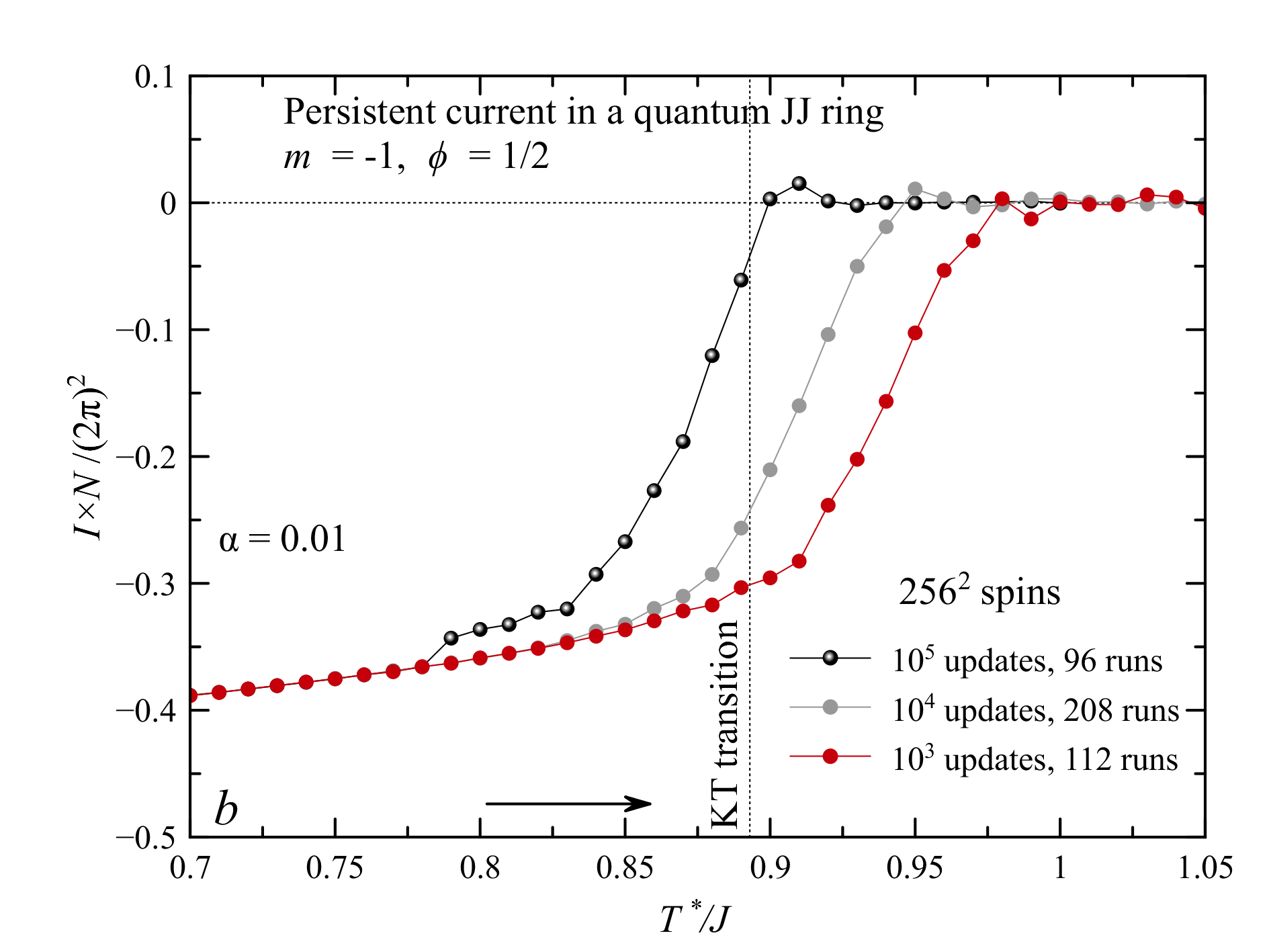}

\caption{Color online: Dependence of the persistent current on $T^{*}$ in
the half-fluxon case. ($a$) For different sizes with low damping
($\alpha=0.01$) and large number of updates; ($b$) For different
numbers of updates.}

\label{Fig:I_vs_T_half-fluxon}
\end{figure}

Results of computations for the system of size $1000^{2}$ in the
half-fluxon case, performed with the intermediate damping $\alpha=0.1$
and a moderate number 3000 of updates for each temperature increased
in small steps (without pre-thermalization with $\alpha=1$), are
shown in Fig. \ref{Fig:I_m_vs_T_Nx=00003D00003DNy=00003D00003D100}.
At the first sight $I(T^{*})$ looks like the temperature dependence
of the order parameter that for the biggest system size disappears
at $T^{*}/J\approx0.97$. The numerical data at low temperatures are
very smooth and precise, while the agreement with the spin-wave theory,
Eq. (\ref{eq:I_SWT}), is excellent. However, there are considerable
fluctuations in a wide critical region, even for a large system size.
They are related to the smallness of the current, $I\propto1/N$.
The average phase change $\left\langle m\right\rangle $ of Eq. (\ref{sum-1})
is exponentially close to an integer set as the initial condition
at low temperatures. This justifies the approximation made in the
derivation of Eq. (\ref{eq:I_SWT}). Temperature dependence of $\left\langle m\right\rangle $
is due to the creation and unbinding of vortex-antivortex pairs, as
explained above.

Computations performed with low damping $\alpha=0.01$ (making pre-thermalization
with $\alpha=1$) and large number of updates (up to $10^{5}$) show
instability of the persistent current in the half-fluxon case, related
to its tunneling between the two equal-energy classical states. In
Fig. \ref{Fig:I_vs_T_half-fluxon}$a$ one can see that for small
system sizes $I$ is getting destroyed by the jumping to the same-energy
state well below the Kosterlitz-Thouless transition temperature. With
increasing the size, the curves $I(T^{*})$ are shifting to the right
and saturate at size $142^{2}$. This behavior will be explained qualitatively
in the next section. Incidentally, for this size and also for $N=256^{2}$
the current is disappearing exactly at $T_{KT}$. This has to be taken
with a grain of salt, however. The curves for size $256^{2}$ in Fig.
\ref{Fig:I_vs_T_half-fluxon}$b$, obtained with temperature increasing,
are shifting to the left when more computer time is allowed for relaxation,
that is, with the number of updates. Although it would take impractically
long computer time to increase the number of updates past $10^{5}$,
one can project that for a sufficient number of updates the persistent
current will disappear at any temperature.

\section{Analysis}

\label{Sec:Analysis}

Here we estimate the energy barrier for the quantum phase slip associated
with the unbinding of vortex-antivortex pairs in a system of finite
size. The energy of a vortex pair has the form
\begin{equation}
E(r)=2E_{c}+2\pi\rho(T)J\ln\left(r/a\right),
\end{equation}
where $E_{c}\sim J$ is the vortex core energy, $a$ is the lattice
spacing, $r$ is the distance between the singularities in the pair,
and $\rho(T)$ is helicity that describes the softening of the system
with increasing temperature, $\rho(0)=1$. Since the energy of the
pair increases with the distance, the unbinding takes place in one
of two cases: 1) $r$ reaches the linear size of the system $L=a\sqrt{N}$;
2) $r$ reaches $r^{*}=a/\sqrt{f_{V}}$, the average distance between
singularities. In the latter case realized for $r^{*}\lesssim L$,
vortices and antivortices, after reaching this distance, recombine
with the members of other pairs, which facilitates the phase slips.
The vorticity and $r^{*}$ can be estimated as
\begin{equation}
f_{V}=e^{-2E_{c}/T},\qquad r^{*}=ae^{E_{c}/T}.
\end{equation}

For $r^{*}\lesssim L$ the barrier $\Delta E$ for the unbinding is
given by $E(r^{*})$,
\begin{equation}
\Delta E=E(r^{*})=2E_{c}+\frac{2\pi\rho(T)JE_{c}}{T},
\end{equation}
where the second term is dominant at lower temperatures. One can see
that $\Delta E$ is independent of the system size. In this case the
rate of the vortex-pair unbinding and, thus, the probability of the
phase slip, are proportional to
\begin{equation}
\exp\left(-\frac{\Delta E}{T}\right)\cong\exp\left(-\frac{2\pi\rho(T)JE_{c}}{T^{2}}\right),\label{eq:Unbinding_rate_large_size}
\end{equation}
that vanishes quickly at low temperatures. This explains why it is
so difficult to obtain numerically the equilibrium result $I=0$ in
the half-fluxon case.

For systems of smaller size at low temperatures there are too few
vortices, so that $r^{*}>L$ and the barrier is given by
\begin{equation}
\Delta E=2E_{c}+2\pi\rho(T)J\ln\left(L/a\right).
\end{equation}
With the second term dominant, this leads to the phase slip (unbinding)
rate proportional to
\begin{equation}
\exp\left(-\frac{\Delta E}{T}\right)\cong\left(\frac{a}{L}\right)^{2\pi\rho(T)J/T}=\frac{1}{N^{\pi\rho(T)J/T}},\label{eq:Unbinding_rate_small_size}
\end{equation}
which desreases with the system size. One can see in Fig. \ref{Fig:I_vs_T_half-fluxon}$a$
that for small sizes $I(T^{*})$ is shifting to the right in accordance
with Eq. (\ref{eq:Unbinding_rate_small_size}) and then saturates
in accordance with Eq. (\ref{eq:Unbinding_rate_large_size}).

\section{Discussion}

\label{Sec:Discussion}

We have studied the dependence of the persistent current in a Josephson
junction ring on the strength of quantum fluctuations, $U/J$, the
number of superconducting islands, $N$, and the flux shreading the
ring, $\phi=\Phi/\Phi_{0}$. All three parameters can be controlled
in experiment. The strength of quantum fluctuations is determined
by the Josephson coupling of the islands and their capacitances. At
some critical $U/J$ the system undergoes a quantum phase transition
into the superinsulator state characterized by zero conductivity of
the ring in the presence of the superconductivity of the islands.
Persistent current presents a good experimental tool for the investigation
of this transition.

From theoretical perspective the problem allows accurate numerical
studies by Monte Carlo techniques developed for spin systems because
it maps onto a classical $2d$ $xy$ model at finite temperature.
In this approach the critical strength of quantum fluctuations projects
onto the temperature of the Kosterlitz-Thouless phase transition.
The numerical solution of the problem is, however, significantly more
challenging than equilibrium problems of spin physics because the
persistent current disappears in the thermodynamic limit $N\rightarrow\infty$.

In accordance with the expectation we find that in all cases the persistent
current is destroyed by sufficiently strong quantum fluctuations.
The manner in which it is destroyed depends strongly on the number
of superconducting islands in the Josephson junction ring and on the
magnetic flux threading the ring. In cases of a half-integer flux,
$\phi=n+1/2$, when the classical ground state of the ring is degenerate,
quantum fluctuations of any strength in theory destroy the persistent
current. However, the phase slips required for that have a finite
probability that is exponentially small at $U\ll J$. In that sense,
the persistent current in the presence of weak quantum fluctuations
is as stable as the magnetic moment of a superparamagnetic particle
below the blocking temperature. Unlike the magnetization of a superparamagnetic
particle (or of a finite-size Ising system) though, the persistent
current cannot be made stable by increasing the system size. This
is because the barrier for the switching of the current due to the
quantum phase slip becomes size independent for large sizes. For small
Josephson rings, we find that quantum fluctuations destroy the persistent
current even faster. However, the dependence on the size is at best
logarithmic. This finding can be of interest to the experimentalists.

When the classical ground state is not degenerate, the persistent
current has a non-zero equilibrium value. At $U\ll J$ the spin-wave
approximation provides an excellent agreement with the numerical data
on how the persistent current decreases when the strength of quantum
fluctuations increases. The behavior is independent of the size of
the ring. This prediction of the theory would be interesting to test
in a real experiment. Another interesting prediction is that the departure
from the universal behavior towards the behavior that depends on the
size of the ring begins at the critical strength of quantum fluctuations
determined by the bulk Kosterlitz-Thouless temperature. We present
numerical study of the phase slips due to the unbinding of vortex-antivortex
pairs and analytical arguments that explain the size-dependence of
the persistent current.

\section{Acknowledgements}

This work has been supported by the grant No. DE-FG02-93ER45487 funded
by the U.S. Department of Energy, Office of Science.


\begin{thebibliography}{10}
\bibitem{Girvin} See, e.g., S. L. Sondhi, S. M. Girvin, J. P. Carini,
and D. Shahar, Continuous quantum phase transitions, Review of Modern
Physics \textbf{69}, 315-333 (1997).

\bibitem{Sachdev} S. Sachdev, \textit{Quantum Phase Transitions}
(Cambridge University Press, Cambridge, UK, 2011).

\bibitem{Ioffe} L. B. Ioffe, M. V. Feigel'man, A. Ioselevich, D.
Ivanov, M. Troyer, and G. Blatter, Topologically protected quantum
bits using Josephson junction arrays, Nature \textbf{415}, 503-506
(2002).

\bibitem{Gladchenko} S. Gladchenko, D. Olaya, E. Dupont-Ferrier,
B. Dou\,{c}ot, L. B. Ioffe, and M. E. Gershenson, Superconducting
Nanocircuits for Topologically Protected Qubits, Nature Physics \textbf{5},
48-53 (2009).

\bibitem{Manucharyan} V. E. Manucharyan, J. Koch, L. I. Glazman,
and M. H. Devoret, Fluxonium: Single Cooper-Pair Circuit Free of Charge
Offsets, Science \textbf{326}, 113-116 (2009).

\bibitem{Zaikin} A. D. Zaikin, D. S. Golubev, A. van Otterlo, and
G. T. Zimanyi, Quantum phase slips and transport in ultrathin superconducting
wires, Physical Review Letters \textbf{78}, 1552-1555 (1997).

\bibitem{Golubev} D. S. Golubev and A. D. Zaikin, Quantum tunneling
of the order parameter in superconducting nanowires, Physical Review
B \textbf{64}, 014504-(14) (2001).

\bibitem{Rastelli} G. Rastelli, I. M. Pop, and F. W . J. Hekking,
Quantum phase-slips in Josephson junction rings, Physical Review B
\textbf{87}, 174513-(18) (2013).

\bibitem{Bradley} R. M. Bradley and S. Doniach, Quantum fluctuations
in chains of Josephson junctions, Physical Review B \textbf{30}, 1138-1147
(1984).

\bibitem{Korshunov} S. E. Korshunov, Effect of dissipation on the
low-temperature properties of a tunnel-junction chain, Soviet Physics
JETP \textbf{68}, 609-618 (1989).

\bibitem{Chow} E. Chow, P. Delsing, and D. B. Haviland, Length-scale
dependence of the superconductor-to-insulator quantum phase transition
in one dimension. Physical Review Letters \textbf{81}, 204-207 (1998).

\bibitem{Mooij-2015} See review and references therein: J. E. Mooij,
G. Schön, A. Shnirman, T. Fuse, C. J. P. M. Harmans, H. Rotzinger,
and A. H. Verbruggen, Superconductor-insulator transition in nanowires
and nanowire arrays, New Journal of Physics \textbf{17}, 033006-(12)
(2015).

\bibitem{Oreg} See, e.g., G. Schwiete and Y. Oreg, Persistent current
in small superconducting rings, Physical Review Letters \textbf{103},
037001-(4) (2009), and references therein.

\bibitem{Choi-1993} M. Y. Choi, Persistent current and voltage in
a ring of Josephson junctions, Physical Review B \textbf{48}, 15920-15925
(1993).

\bibitem{Matveev} K. A. Matveev, A. I. Larkin, and L. I. Glazman,
Persistent current in superconducting nanorings, Physical Review Letters
\textbf{89}, 096802-(4) (2002).

\bibitem{Lee-2003} M. Lee, M.-S. Choi, and M. Y. Choi, Quantum phase
transitions and persistent currents in Josephson-junction ladders,
Physical Review B \textbf{68}, 144506-(11) (2003).

\bibitem{Pop} I. M. Pop, I. Protopopov, F. Lecocq, Z. Peng, B. Pannetier,
O. Buisson, and W. Guichard, Measurement of the effect of quantum
phase-slips in a Josephson junction chain, Nature Physics \textbf{6},
589-592 (2010).

\bibitem{Tinkham} M. Tinkham, \textit{Introduction to Superconductivity}
(Dover Publications, 2004, ISBN 0-486-43503-2).

\bibitem{Tobochnik} J. Tobochnik and G. V. Chester, Monte Carlo study
of the planar spin model, Physical Review B \textbf{20}, 3761-3769
(1979).

\bibitem{Gupta} R. Gupta and C. F. Baillie, Critical behavior of
the two-dimensional XY model, Physical Review B \textbf{45}, 2883-2898
(1992).

\bibitem{Freedman} J. R. Friedman, V. Patel, W. Chen, S. K. Tolpygo,
and J. E. Lukens, Quantum superposition of distinct macroscopic states,
Nature \textbf{406}, 43-45 (2000).

\bibitem{Mooij} C. H. van der Val, A. C. ter Haar, F. K. Wilhelm,
R. N. Schouten, C. J. Harmans, T. P. Orlando, S. Lloyd, and J. E.
Mooij, Quantum superposition of macroscopic persistent-current states,
Science \textbf{290}, 773-777 (2000). \end{thebibliography}
\end{document}